\documentclass[twocolumn]{aastex61}

\usepackage{amsmath}
\usepackage{color}

\begin{document}

\hyphenation{analy-sis}
\hyphenation{analy-se}


\newcommand{\avdt}{\langle|\delta\theta|\rangle}

\title{The uncertainty of local background magnetic field orientation in anisotropic plasma turbulence}

\correspondingauthor{Felix Gerick}
\email{felix.gerick@uni-koeln.de}

\author[0000-0001-9924-0562]{F. Gerick}

\affil{Institute of Geophysics and Meteorology, University of Cologne, Cologne, Germany}

\author[0000-0003-1413-1231]{J. Saur}
\affil{Institute of Geophysics and Meteorology, University of Cologne, Cologne, Germany}
\author[0000-0001-5030-1643]{M. von Papen}
\affil{Institute of Geophysics and Meteorology, University of Cologne, Cologne, Germany}

\begin{abstract}
In order to resolve and characterize anisotropy in turbulent plasma flows a proper estimation of the background magnetic field is crucially important. Various approaches to calculate the background magnetic fields, ranging from local fields to globally averaged fields, are commonly used in the analysis of turbulent data. Here we investigate how the uncertainty in the orientation of a scale dependent background magnetic field influences the ability to resolve anisotropy. Therefore we introduce a quantitative measure, the \emph{angle uncertainty}, which characterizes the uncertainty of the orientation of the background magnetic field which turbulent structures are exposed to. The angle uncertainty can be used as a condition to estimate the ability to resolve anisotropy with certain accuracy. We apply our description to resolve spectral anisotropy in fast solar wind data. We show that if the angle uncertainty grows too large, the power of the turbulent fluctuations is attributed to false local magnetic field angles, which may lead to an incorrect estimation of spectral indices. In our results an apparent robustness of the spectral anisotropy to false local magnetic field angles is observed, which can be explained by a stronger increase of power for lower frequencies when the scale of the local magnetic field is increased. The frequency dependent angle uncertainty is a measure which can be applied to any turbulent system. 
\end{abstract}


\renewcommand{\vec}{\mathbf}
\newcommand{\rmsdt}{\mathrm{RMS}(\delta\theta)}
\newcommand{\sbmin}{s_b(\sigma)}

\section{Introduction} \label{sec:intro}

Turbulent flows in magnetized plasmas are anisotropic due to the presence of a magnetic field (see, e.g. reviews by \citealt{Horbury2012} and \citealt{Oughton2015}). In contrast to the velocity field, for the magnetic field no Galileo transformation exists such that for a certain eddy (or turbulent structure), the magnetic field associated with larger eddies vanishes. Therefore the magnetic field of all larger scales directly influences the smaller scales of turbulence.

Deciphering the anisotropic structure of plasma turbulence is a major challenge and several models are debated in the literature (e.g., \citealt{Matthaeus1990, GS95, Bieber1996, Saur1999, Galtier2005, Boldyrev2006, Galtier2006, Beresnyak2008, Howes2008, Howes2011, Boldyrev2012, Narita2015}). For understanding the anisotropy of turbulence in magnetized plasmas, the spatial and temporal extents of the magnetic field controlling the orientation and the decay of the turbulent eddies of specific scales remains unclear, but is of crucial importance.

Two approaches are commonly used to characterize the controlling scale of the magnetic field, referred to as global and local frame \citep{Maron2001, Horbury2008, Beresnyak2009a, Cho2009, Tessein2009, Chen2011, Matthaeus2012}. In the global frame the magnetic field $\vec{B}(t)$ is averaged over scales much larger than the correlation length of the turbulent fluctuations to obtain the global mean field $\vec{B}_0$. In the local frame, on the other hand, it is assumed that a magnetic field at scales on the same order as those given by the individual turbulent structure or eddy control the anisotropy of the turbulence \citep{Cho2000,Maron2001,Cho2002,Cho2004}.

These considerations on the controlling scales are relevant for magnetized plasmas whether observed in space or generated in numerical simulations. They impose  important questions if the turbulent flow contains fluctuations $\vec{\delta b}{=}\vec{B}-\vec{B}_0$ with a root mean square (RMS) similar or larger compared to the mean magnetic field obtained by averaging over global scales. Only in the case of $\vec{B}_0{\gg} \vec{\delta b}$, the problem simplifies because the local background field is approximately equal to the global mean field.

The solar wind is a  medium where the large scale background magnetic field $\vec{B}_0$, averaged over hours, days or years is often on the same order as the RMS of the magnetic field fluctuations $\vec{\delta b}$.
Solar wind studies using a global magnetic field frame only detected anisotropy in the power of the fluctuations, but no anisotropy in the spectral index \citep{Tessein2009}.
On the other hand several studies using a local and scale dependent magnetic field for the analysis have revealed anisotropy in both power and spectral index $\kappa$ in the inertial range spectrum of solar wind data \citep{Horbury2008, Alexandrova2008, Podesta2009, Chen2010, Luo2010, Wicks2010, Wicks2011, Podesta2013}. \citet{Horbury2008}, for the first time, analyzed the spectral index $\kappa$ with respect to a background magnetic field using such a local frame. The observed spectrum showed a spectral index of -2 parallel compared to -5/3 perpendicular to the local magnetic field (see Figure 2, lower panel, in \citealt{Horbury2008}), which is in agreement with the predicted scalings of the critical-balance theory \citep{GS95}. Besides the spectral index, other anisotropic properties have been successfully analyzed using a local and scale dependent magnetic field (e.g., \citealt{Salem2012,He2013,Bruno2015}).

Here we introduce a necessary geometrical condition for the scales of the magnetic field to observationally resolve anisotropy within measured or simulated data. This condition is given by the average uncertainty in the orientation of the local background magnetic field at a certain scale. This uncertainty in the orientation is measured by the angle of the local background magnetic field with respect to the orientation of the observed fluctuations. This is referred to as \emph{angle uncertainty} in the remainder of this work. To quantify the angle uncertainty in spacecraft measurements, we use the \emph{field to flow angle} $\theta$, which is defined as the angle between the magnetic field $\vec{B}$ and the unperturbed flow direction of the solar wind $\vec{v}_\mathrm{SW}$. In other systems the orientation of the field may be conveniently defined in a different way.

\begin{figure}
\includegraphics[width=\linewidth]{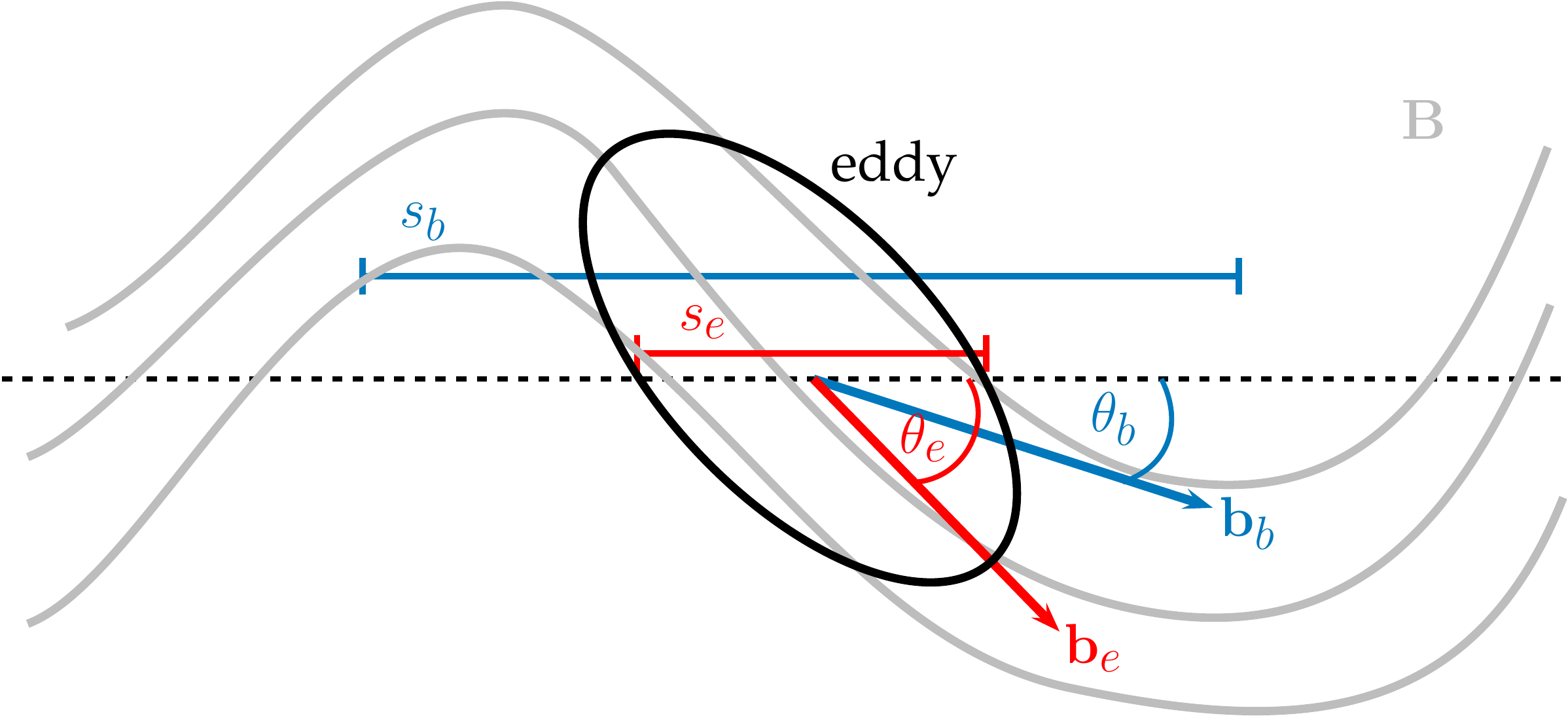}
\caption{\label{fig:intro}Schematic representation of the orientation of an eddy with respect to the orientation of two different background fields averaged at the scale of the eddy $s_e$ and averaged at some larger scale $s_b$.}
\end{figure}

The orientation of an elongated eddy within a magnetic vector field $\vec{B}$ is shown schematically in Figure \ref{fig:intro}. Measuring along the dashed line the magnetic field averaged over scale $s_e$, which characterizes the size of an eddy (detailed definition in the following section), is associated with the angle $\theta_e$. 
If the associated background magnetic field is defined over a larger scale $s_b$ the field to flow angle is $\theta_b$. The fluctuations observed at the eddy scale are in this case associated with a different field to flow angle. We hypothesize that if the angle discrepancy between the scale at which the magnetic field is averaged and the eddy scale grows beyond a certain threshold, the angle of the local magnetic field is no longer well estimated. Therefore, the anisotropic properties of turbulent eddies might not be resolved under the assumption that the orientation of the eddies adjust locally to the magnetic field. 

In the following we define the necessary scales, give a mathematical definition of the background magnetic field for different levels of localizations, and formally introduce the angle uncertainty as a measure for the orientation of an eddy within such averaged magnetic fields. Subsequently we apply it to 91 days of magnetic field measurements within the fast solar wind \citep{Wicks2010} and explore its suitability as a necessary condition to resolve observed or expected solar wind spectral anisotropy.

\section{Analysis of spectral anisotropy} \label{sec:methods}
\subsection{Relevant Scales \& Wavelet Method} \label{ch:wavelets}

To analyze spectral properties of turbulent fluctuations we use a method based on the wavelet transformation. We denote $B_i(t)$ with $i{=}R,T,N$ the magnetic field components measured as a function of time $t$ in the RTN-coordinate system\footnote{The unit vector $\vec{e}_R$ points radially away from the sun, $\vec{e}_T{=}\vec{e}_\Omega\times\vec{e}_R$ is perpendicular to $\vec{e}_R$ and the sun's rotational axis $\vec{e}_\Omega$ and $\vec{e}_N{=}\vec{e}_R\times\vec{e}_T$ completes the right handed system.}.

\begin{figure}
\centering
\includegraphics[width=\linewidth]{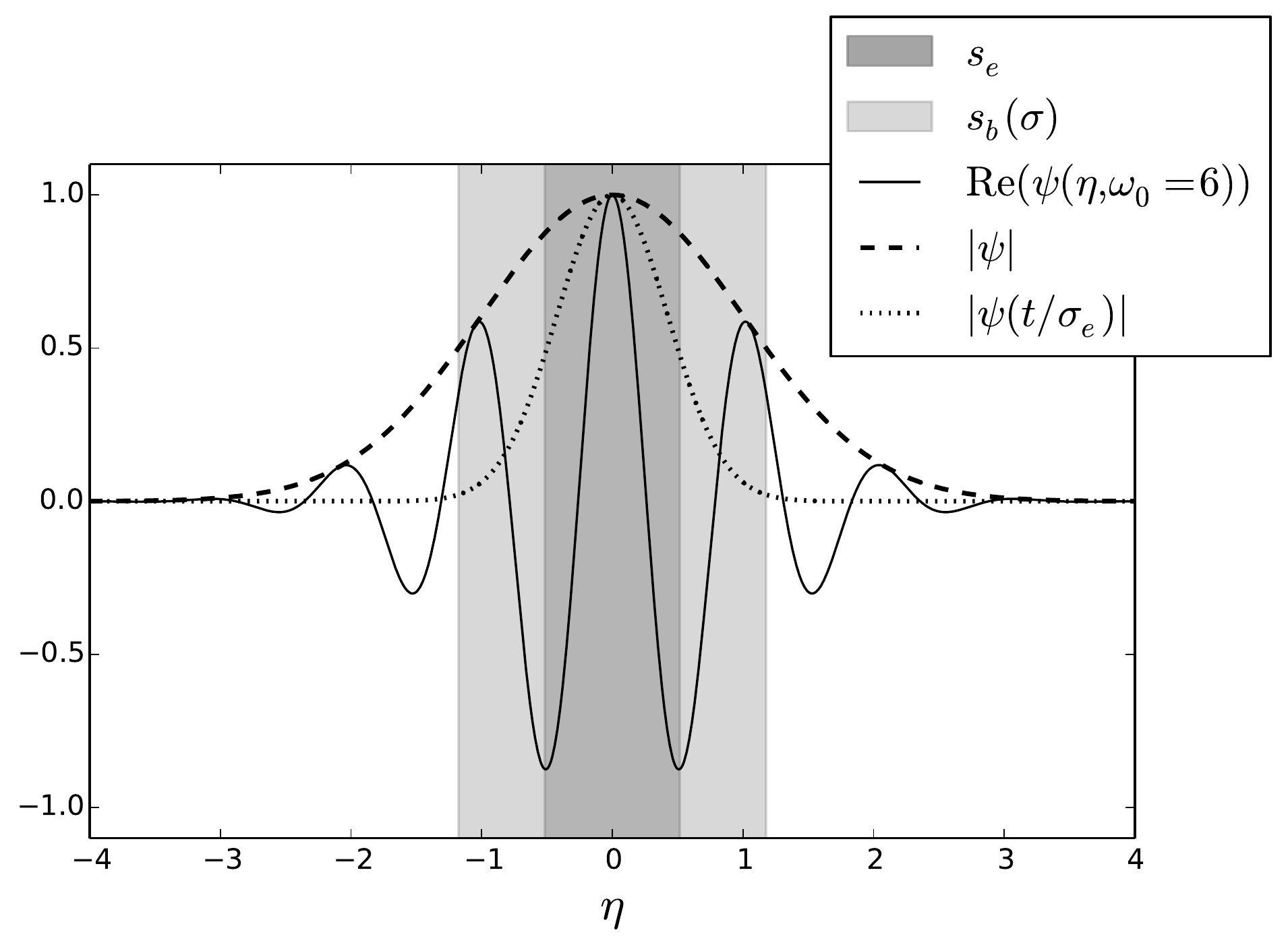}
\caption{Real part of the normalized Morlet wavelet in time domain with the amplitude envelope (dashed), the Gaussian with standard deviation $\sigma_e$ and the two characteristic time scales $s_e$ (dark grey area) and $s_b(\sigma)$ (light grey area).}
\label{fig:morlet}
\end{figure}

The wavelet transformation of the components $B_i(t)$ is calculated as
\begin{equation}
W_i(t,\sigma)=\frac{1}{\sqrt{\sigma}}\int\limits B_i(t')\psi\left(\frac{t'-t}{\sigma}\right)dt',
\label{eq:wavelet}
\end{equation}
where $\psi(\eta)$ is the mother wavelet and $\sigma$ the wavelet scale. 
The absolute squared values of the complex wavelet coefficients, $|W_i(t,\sigma)|^2$, give energy density at time $t$ and wavelet scale $\sigma$. 
In case of the Morlet wavelet 
\begin{equation}
\psi(\eta)=\pi^{-1/4}e^{-i\omega_0 \eta}e^{-\eta^2/2}
\label{eq:morlet}
\end{equation}
the wavelet scale is the standard deviation of the Gaussian amplitude envelope of the wavelet displayed as dashed line in Figure \ref{fig:morlet} \citep{TC98}. The width of the wavelet can thus be defined by the full width at half maximum of the Gaussian window $2\sqrt{2\ln(2)}\sigma$ (shown as light grey area in Figure \ref{fig:morlet}). 
The wavelet packet given by Equation \eqref{eq:morlet} can be associated with two scales. One scale $s_e$ is associated with the period (frequency) of the fluctuations of the turbulent eddy that is to be analyzed. The other scale $s_b$ is associated with the full width at half maximum of the wavelet, which constrains the temporal resolution and which is used in the following section for definition of the local background magnetic field.

The translation from wavelet scale to frequency 
\begin{equation}
f_e=(\omega_0{+}\sqrt{2{+}\omega_0^2})/(4\pi\sigma)
\end{equation}
depends on the number of oscillations $\omega_0$ within the wavelet \citep{Meyers1993}. Here we use $\omega_0{=}6$ so that $f_e{=}(1.033\sigma)^{-1}$. We define the scale $s_e$ of the eddy under consideration as one period of the frequency $1/f_e$ (shown in Figure \ref{fig:morlet}, dark grey area) independent of the choice of $\omega_0$. The energy density of the wavelet coefficients can be associated with the eddy frequency $f_e$, which is consistent with the classical Fourier analysis commonly used in turbulence analysis.

To associate a local background magnetic field to each wavelet coefficient one can use a Gaussian with standard deviation $\sigma_b{=}\sigma$. This is a reasonable choice as it describes the scale over which the energy density in the wavelet is calculated \citep{Horbury2008}. In the remainder of this work we also investigate background magnetic fields averaged over larger scales using Gaussian windows with standard deviation $\sigma_b {>} \sigma$. We therefore use a local, scale dependent background magnetic field $b_i$ for each component $i$ given by

\begin{equation}
b_i(t,s_b) = \int B_i(t') \exp\left(\frac{-(t'-t)^2}{2\sigma_b^2}\right) dt',\label{eq:magsmooth}
\end{equation}
the convolution of the magnetic field with a Gaussian \citep{Horbury2008, Podesta2009}. We introduce a dimensionless factor $\alpha$, so that $\sigma_b{=}\alpha \sigma$, to quantify the increase of the averaging width. In case of $\alpha{=}1$ the averaging width corresponds to the envelope of the wavelet. The total averaging scale is $s_b(\sigma_b){=}2\sqrt{2\ln(2)}\sigma_b$. Standard deviations smaller than $\sigma_b{=}\sigma$ should not be used to average the magnetic field since the energy density of the associated wavelet coefficients would correspond to wavelets larger than the averaged magnetic field.

The ratio of the smallest possible averaging scale and the eddy scale 
\begin{equation}
\frac{s_b(\sigma)}{s_e}=\frac{2\sqrt{2\ln{2}}(\omega_0+\sqrt{2+\omega_0^2})}{4\pi},\label{eq:se-sb-ratio}
\end{equation} 
depends only on the choice of $\omega_0$. This ratio is always larger than one and increases with larger $\omega_0$. The most local choice would be $\omega_0{=}6$ as $\omega_0{<}6$ fails the admissibility condition of wavelets \citep{Farge1992}. That is why there is a minimum difference between $s_e$ and $s_b$ for wavelet based analysis. For $\omega_0{=}6$ the minimum averaging scale $s_b(\sigma)$ is $2.28$ times larger than $s_e$.

\subsection{Field to Flow Angles \& Uncertainty}
 
To compute a field to flow angle
\begin{equation}
\theta(t,s_b)=\cos^{-1}\left(\frac{\vec{b}(t,s_b)\cdot \vec{v}_\mathrm{sw}}{|\vec{b}(t,s_b)||\vec{v}_\mathrm{sw}|}\right),
\label{eq:theta}
\end{equation}
one can use the local background magnetic field vector $\vec{b}(t,s_b)$ obtained from equation \eqref{eq:magsmooth}. This gives the angle between the (local) background magnetic field to the average solar wind velocity $\vec{v}_\mathrm{sw}$. The second angle which describes the orientation of the local background magnetic field is the azimuth angle, but studies have shown that spectral anisotropy is approximately azimuthally symmetric around the local background magnetic field \citep{Horbury2008,Podesta2009}. We therefore only consider $\theta$ to characterize the variability of the orientation of the magnetic field.

The global power spectral density (PSD) at a distinct field to flow angle and with a temporal resolution $\Delta t$ can be obtained from the wavelet coefficients by
\begin{equation}
P(f_e; \theta)=\sum_{i=R,T,N} P_i(f_e; \theta),
\end{equation}
where
\begin{equation}
P_i(f_e; \theta)=\frac{2\Delta t}{N}\sum_{j=1}^N\left|W_i(t_j,\sigma;\theta)\right|^2 
\label{eq:psd}
\end{equation}
is computed from $N$ wavelet coefficients $W_i(t_j,\sigma;\theta)$ associated with the angle $\theta(t_j,s_b)$. In our analysis we calculate average $P(f_e; \theta)$ within bins of $\theta = 0{-}10^\circ, 10{-}20^\circ, \dots ,80 {-} 90^\circ$.

We averaged the magnetic field time series according to Equation \eqref{eq:magsmooth} and calculated scale dependent angles that characterize the orientation of an associated local background magnetic field according to Equation \eqref{eq:theta}. We now define the \emph{angle uncertainty}
\begin{equation}
\delta\theta(t,s_b,s_e)=\theta(t,s_e)-\theta(t,s_b). \label{eq:deltatheta01}
\end{equation}
This angle quantifies the difference in magnetic field orientation between the eddy of scale $s_e$ and the (larger) scale $s_b$ over which the local magnetic field is defined. 

We hypothesize that this newly introduced quantity is an indication for how local an averaged magnetic field is. The root mean square of the angle uncertainty $\delta\theta$ can be used to describe the average uncertainty of the orientation of eddies of size $s_e$ when their orientation is measured with respect to a larger local background magnetic field. A minimum uncertainty arises from the difference between the background magnetic field averaged at scale $s_b$ and the magnetic field averaged with a Gaussian of standard deviation $\sigma_e{=}(2\sqrt{2\ln(2)})^{-1}s_e$, which corresponds to the scale of the eddy fluctuations $s_e$ (shown in Figure \ref{fig:morlet} as dotted line). The minimum uncertainty is an inevitable consequence of wavelet analysis as energy density at eddy scale $s_e$ is averaged over the width of the wavelet $\sbmin$. The frequency uncertainty of the wavelet transform as an additional factor in the angle uncertainty is neglected as it is found to be insignificant compared to the difference between $s_e$ and $\sbmin$.

In the following we use the root mean square $\rmsdt$ to analyze the influence of the angle uncertainty on plasma turbulence properties, i.e. spectral anisotropy at magnetohydrodynamic (MHD) scales discussed here.

\section{Solar Wind Observations}\label{ch:results}

\subsection{Angle Uncertainty \& Spectral Anisotropy}

\begin{figure}
\centering
\includegraphics[width=\linewidth]{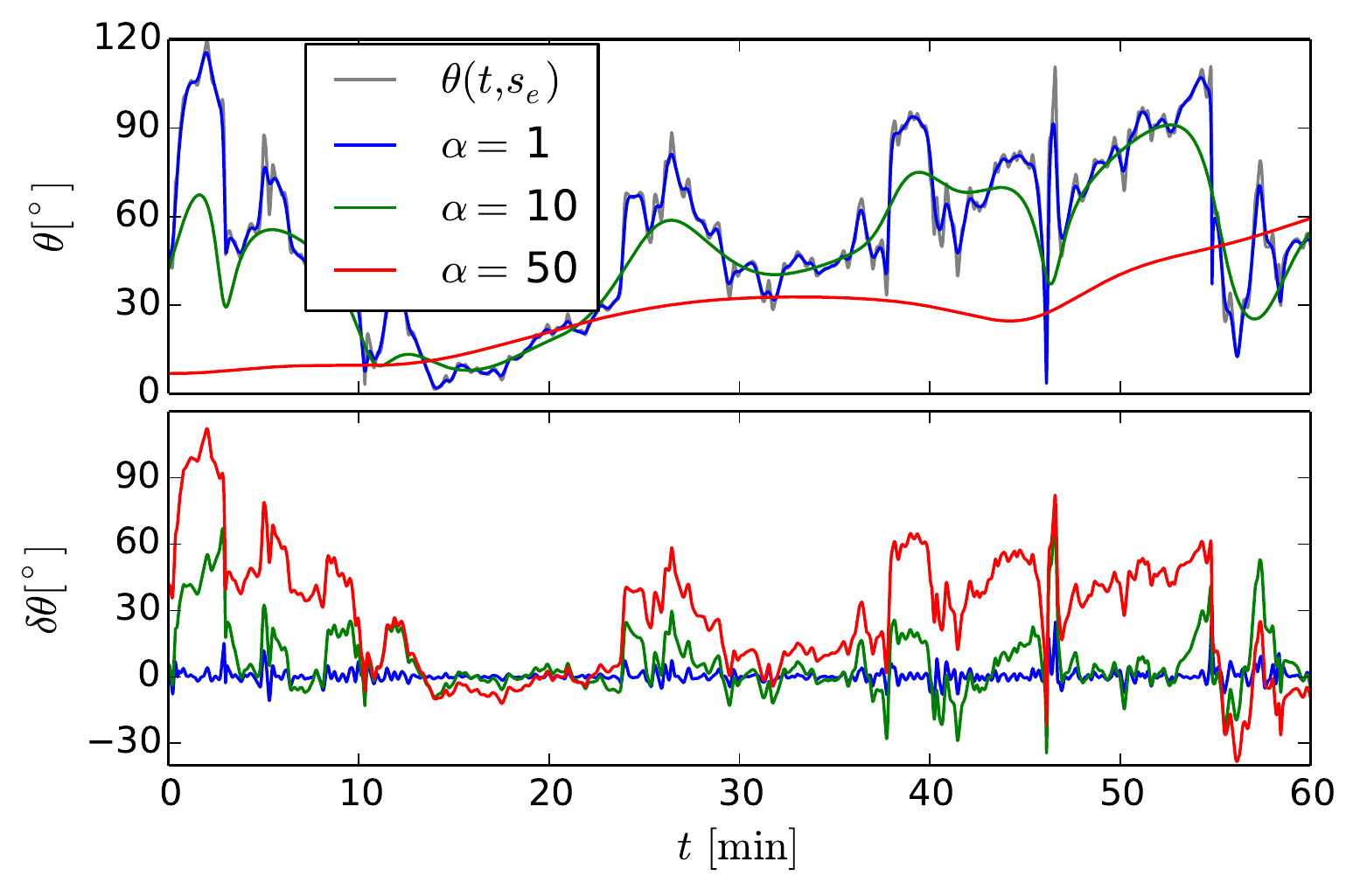}
\caption{$\theta(t,s_b,s_e)$ (top) and $\delta\theta(t,s_b,s_e)$ (bottom) at frequency $f_e{=}0.1$ Hz for different averaging scales $s_b$ of the local background magnetic field in one hour of Ulysses solar wind data (1995, DOY 100, 01:00:55 to 02:00:55).}\label{fig:timeseries}
\end{figure}

We now investigate the scale dependence of the RMS of the angle uncertainty $\delta\theta$ and how it is related to the ability to resolve spectral anisotropy. For that matter we use 91 days of fast solar wind data with a resolution of 1 s from the Ulysses spacecraft from 1995, days 100-190, during a polar orbit at around 1.4-1.9 AU  \citep{Balogh1992,McComas2000,Wicks2010}. Similar and for comparison with  \citet{Horbury2008}, the mean flow velocity of the solar wind is assumed to be in the radial direction. For the the time intervals used in this study, the deviation between the radial direction and the measured solar wind flow is on average 2 degree and can be neglected. We calculate the angle resolved PSD according to Equation \eqref{eq:psd} and the angle uncertainty according to Equation \eqref{eq:deltatheta01} for several different scales on which the average magnetic field is calculated ($\alpha{=}1{-}300$). 

Figure \ref{fig:timeseries} shows an example of $\theta(t,s_b)$ (top) and $\delta\theta(t,s_b,s_e)$ (bottom) at eddy frequency $f_e{=}0.1$ Hz in one hour of the solar wind data. Three different averaging scales $s_b(\alpha\sigma)$ are displayed. For $\alpha{=}1$, $\theta(t,s_b)$ is almost indistinguishable from $\theta(t,s_e)$ and $\delta\theta$ is small (RMS = $3^\circ$). 
The larger the averaging scale $s_b$ represented by the factor $\alpha$, the larger the values of $\delta\theta$. At a factor $\alpha{=}50$, the angle $\theta$ as a function of time becomes fairly smooth compared to the highly fluctuating $\theta(t,s_e)$. The resultant $\delta\theta$ for $\alpha{=}50$ vary strongly and shows values up to $90^\circ$.

\begin{figure}
\centering
\includegraphics[width=\linewidth]{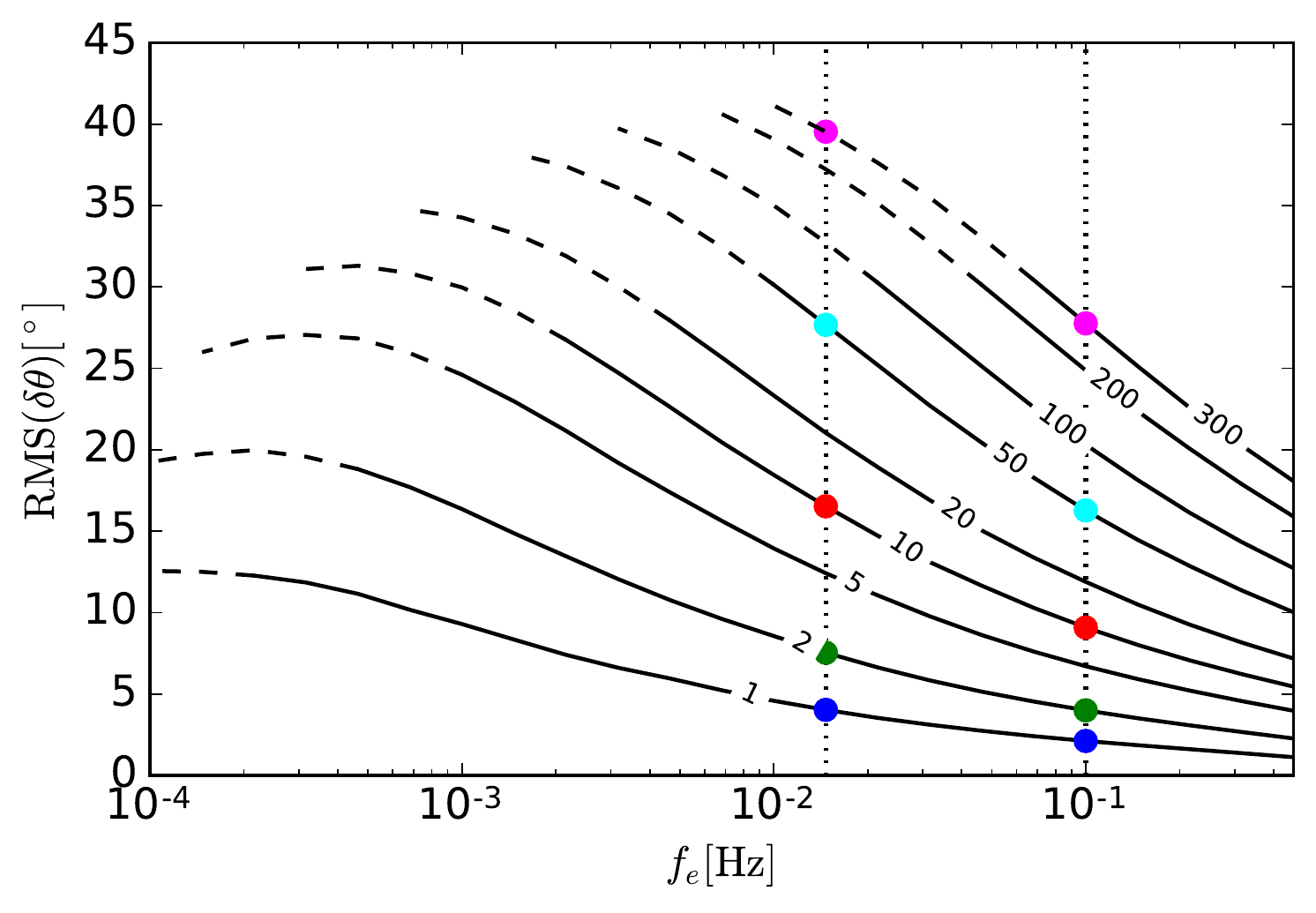}
\caption{RMS of $\delta\theta$ as a function of eddy frequency $f_e$ at different factors $\alpha{=}1, ..., 300$ (indicated as numbers of lines) of the minimum averaging width for characterizing a local background magnetic field. The areas averaged at scales larger than the outer scales $L{\sim}1.5\cdot 10^6$ km \citep{Wicks2010} and $v_\mathrm{sw}{\approx}760$ km/s are indicated by dashed lines. The dotted vertical lines represent the boundary frequencies used for spectral index fitting and the colored dots aid comparison with Figures \ref{fig:psumhisto} and \ref{fig:psds}.}
\label{fig:dtheta_factor}
\end{figure}

The RMS of the angle uncertainty $\delta\theta$ for the complete data set is shown in Figure \ref{fig:dtheta_factor} as a function of eddy frequency $f_e$. We see that the RMS of the angle uncertainty increases with the width of background magnetic field expressed through the factor alpha. This increase is expected from the sample values shown in Figure \ref{fig:timeseries}. $\rmsdt$ also increases as the frequency $f_e$ decreases, which may be explained by the power law increase of power towards lower frequencies in the turbulent cascade. The dotted lines mark the frequency range $15$ mHz ${<}f_e{<}$ 100 mHz considered to be the inertial range and used later to estimate the spectral index \citep{Horbury2008}. In this range $\rmsdt$ reaches values between 7--12$^\circ$ for $\alpha{=}5$. For small factors $\alpha$ around 2-3 the RMS of the angle uncertainty is below 10$^\circ$. At very large factors $\alpha{\geq}50$ the RMS is over 25$^\circ$ and averaging scales reach the length of the outer scale (indicated by dashed lines in Figure \ref{fig:dtheta_factor}) estimated to be $L \sim 6\cdot 10^6$ km by \citet{Wicks2010}. As the power depending on the angle is sorted into $10^\circ$ bins one might expect that the spectral anisotropy vanishes at factors $\alpha{\geq} 5$, because the RMS of the angle uncertainty grows larger than the angle bin. We will analyze this aspect in detail later in this paper. 

\begin{figure}
\centering
\includegraphics[width=\linewidth]{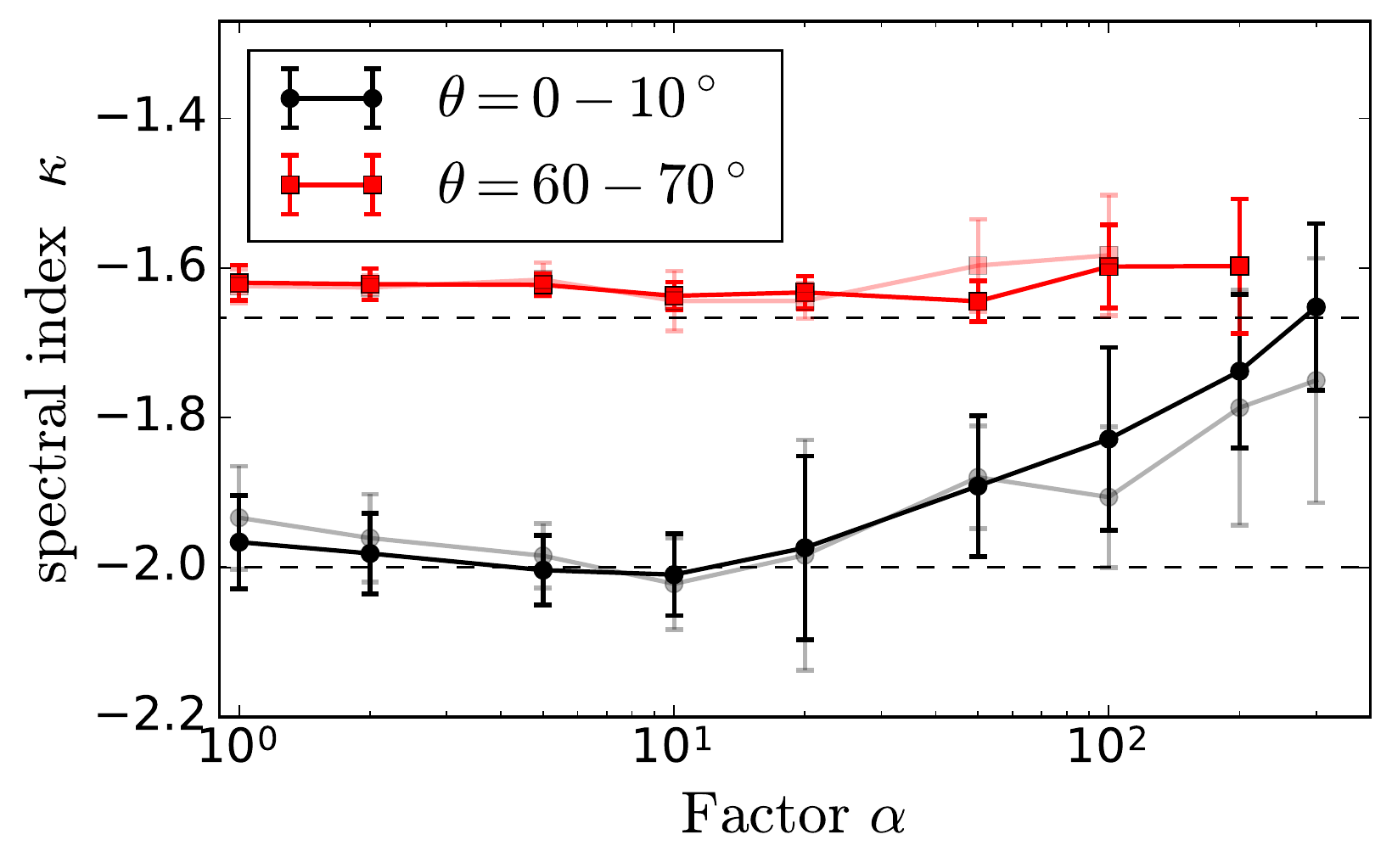}
\caption{Spectral index $\kappa$ at $\theta{=}0{-}10^\circ$ and $\theta{=}60{-}70^\circ$ as a function of increased averaging width by the factor $\alpha$ . Error bars show the 95\% confidence interval of the least squares fit to $P(f_e;\theta)$. Light-colored lines indicate spectral indices from 31 day data (DOY 100-130) of Ulysses data \citep{Horbury2008}. We show angles between $\theta{=}60{-}70^\circ$ as a meaningful average for $\theta{=}80{-}90^\circ$ was not available for large averaging widths.}
\label{fig:correlation_spec_factor}
\end{figure}

For the spectral index analysis we compute the power spectra $P(f_e;\theta)$ and determine the spectral indices in the range of $15 \mathrm{~mHz~}{<}f{<} 100 \mathrm{~mHz}$ for each $\theta(t,s_b(\alpha\sigma))$ bin. The spectral indices at low (0--10$^\circ$) and high (60--70$^\circ$) angles for each factor $\alpha$ are shown in Figure \ref{fig:correlation_spec_factor}. Error bars denote the 95\% confidence interval of the least squares fit in log-space. We show the spectral index at 60--70$^\circ$, because there are not enough coefficients with angles 80--90$^\circ$ to compute a meaningful average. However, the spectral index is approximately constant for angles $\theta{>}50^\circ$ \citep{Horbury2008, Papen2015} and, therefore, the angle bin 60--70$^\circ$ represents the perpendicular case.
For $\alpha{=}1$ the analysis is similar to those of \citet{Horbury2008} and \citet{Podesta2009} and the spectral indices are in agreement with the anisotropic scaling predicted by the critical-balance theory \citep{GS95}. It shows a spectrum $f^{-2}$ parallel and $f^{-5/3}$ perpendicular to the local background magnetic field (see Figure \ref{fig:correlation_spec_factor}, $\alpha{=}1$). We note that the spectral anisotropy is not maximal for $\alpha{=}1$ but for $\alpha{=}10$. However, the difference between the parallel spectral indices corresponding to these factors is small and within error bars. For factors $\alpha{>}10$ the anisotropy slowly decreases. This is also observed in the shorter 31 day data set used by \citet{Horbury2008}, which span DOY 100-130 of year 1995 and is shown in light red and grey lines in Figure \ref{fig:correlation_spec_factor}. The spectral index of the perpendicular cascade is not affected by the scale of the background field. The reason is that for $\alpha{=}1$ the spectral index as a function of the field to flow angle $\theta$ near $0^\circ$ changes very rapidly with growing $\theta$, while it is almost constant at -5/3 for field to flow angles in the range of 30$^\circ$ to 90$^\circ$ (see Figure 2 in \citealt{Horbury2008}). Additionally, the power at smaller angles is sufficiently smaller compared to the power at larger angles, which thus dominate the spectral contributions (again Figure 2 in \citealt{Horbury2008}). The factors $\alpha{>}100$ correspond to a local field so large, namely averaged over $s_b{=}4.5$~h at 15~mHz, that it can be regarded as a global background field. Accordingly, the spectral anisotropy is not resolved for a global field. This is to our knowledge the first time that a gradual change of the spectral anisotropy from local to global field has been shown.

The robustness of the spectral index for factors $5{<}\alpha{<}20$ is unexpected as the previously introduced RMS of the angle uncertainty suggests that anisotropy might not be resolved for factors $\alpha{\geq} 5$, since a strong variability of the  spectral index for $\theta{<}30{-}40^\circ$ is observed and a resolution of $10^\circ$ is necessary (see Figure 2, lower panel, in \citealt{Horbury2008}). To explain this discrepancy between the observed spectral anisotropy and the observed RMS of the angle uncertainty we now study how accurately the wavelet coefficients are associated with the angle bins under consideration.

\subsection{Origin of Power in the Parallel Angle Bin}

In Figure \ref{fig:dtheta_factor} we have shown that large scales of the local background field lead to large angle uncertainties. This can be interpreted in the sense that a large scale local background magnetic field, i.e. a background magnetic field with factors $\alpha\geq5$, is not an adequate representation for the orientation of the turbulent fluctuations. Mathematically, this means that wavelet coefficients $W(t,f_e,\theta)$ are not assigned to the correct angle.

\begin{figure}
\centering
\includegraphics[width=\linewidth]{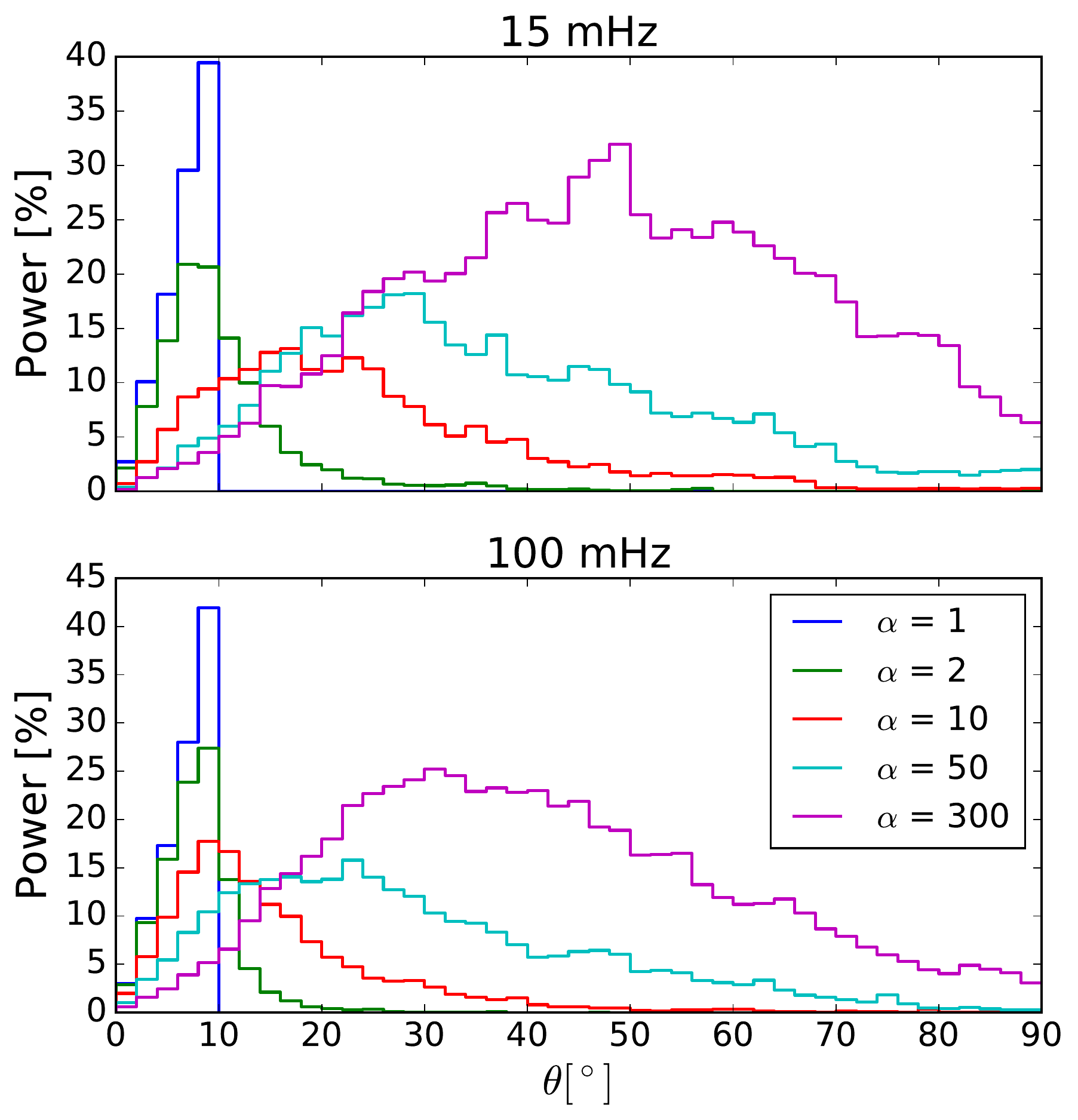}
\caption{Origin $\theta$ of power for wavelet coefficients originally within the $0{-}10^\circ$ bin for $\alpha{=}1$ at 15~mHz (top) and 100~mHz (bottom). Percentages of the power are with respect to the total power in the $0{-}10^\circ$ bin at $\alpha{=}1$. For larger factors $\alpha{>}1$ the power additionaly comes from larger field to flow angles. It can be seen that \emph{false coefficients} first contribute to low frequencies. For $\alpha{=}10$, e.g., the maximum power comes from  $20^\circ$ at 15~mHz but only from $10^\circ$ at 100~mHz.}
\label{fig:psumhisto}
\end{figure}

We now investigate if and how many wavelet coefficients associated with larger angles $\theta(\alpha{=}1){>}10^\circ$ at the most local scale are falsely assigned to the angle bin $0{-}10^\circ$ when the local background field is large ($\alpha{=}2{-}300$). In the following we refer to wavelet coefficients originating from higher angles and being assigned to the $0{-}10^\circ$ bin as \emph{false} coefficients. 

We compute the angular origin $\theta(\alpha{=}1)$ of larger scale ($\alpha{>}1$) coefficients within the $0{-}10^\circ$ bin for the upper (100~mHz) and lower (15~mHz) frequency boundary of the fit range. The result tells us how many \textit{false} coefficients contribute to the power $\sum |W \left( \theta(\alpha) \right) |^2$ within the $0^\circ{\leq}\theta(\alpha){<}10^\circ$ bin and is shown in Figure \ref{fig:psumhisto}. To aid visualization of the redistribution of the angle bins for growing $\alpha$ we choose a bin resolution of $2^\circ$. The histograms are normalized to the total power in the 0--10$^\circ$ bin at $\alpha{=}1$. It can be seen that if we use larger and larger averaging widths, the power spectra include increasingly more false coefficients with angles originally outside the $0{-}10^\circ$ bin. For $\alpha{=}10$ we observe that for 15 mHz the maximum of coefficients actually stems from angles around $20^\circ$ and thus the power has large contributions from \textit{false coefficients}. The corresponding RMS of the angle uncertainties associated with this averaging width $\alpha{=}10$ are RMS$(\delta\theta){>}15^\circ$ (shown in Figure \ref{fig:dtheta_factor}).

\begin{figure}
\centering
\includegraphics[width=\linewidth]{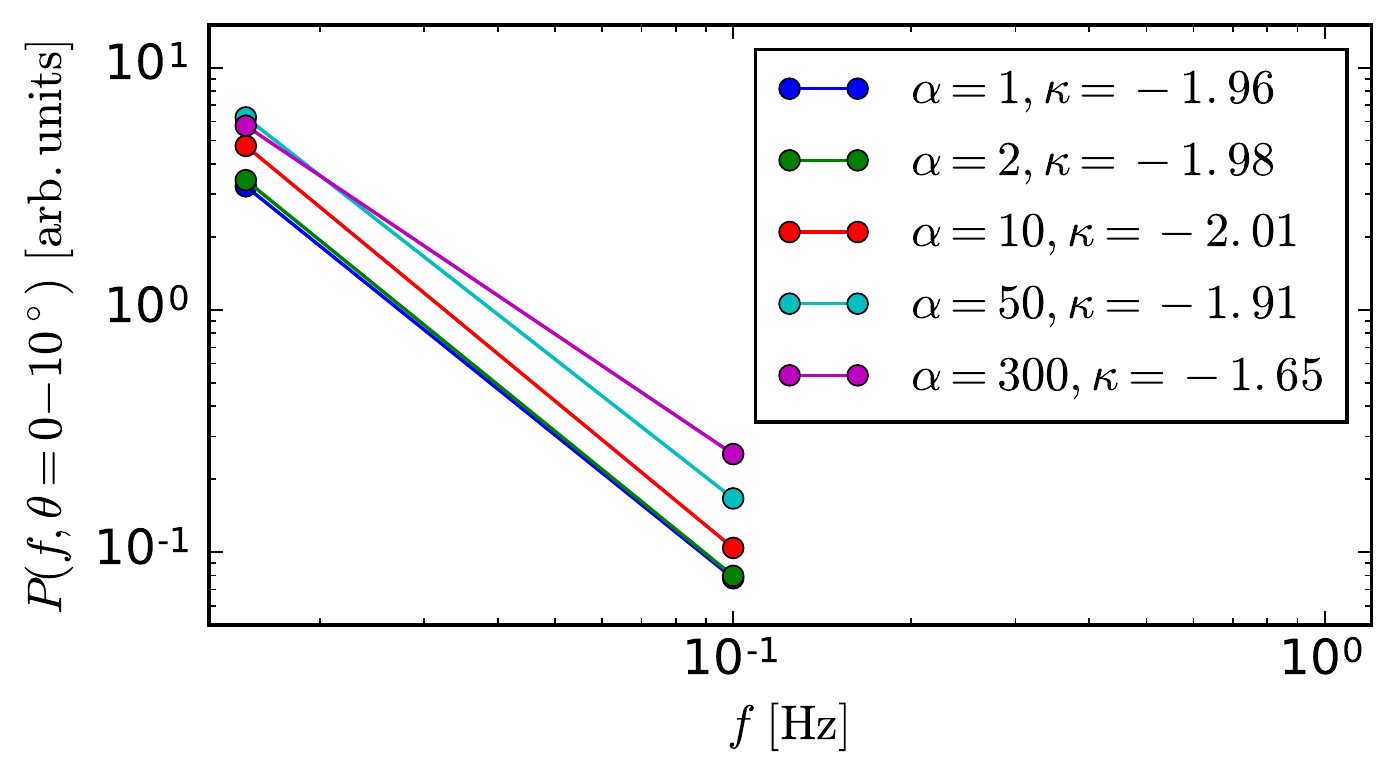}
\caption{PSD and spectral index $\kappa$ at $\theta{=}0{-}10^\circ$ at the factors 1, 2, 10, 50 and 300 in blue, green, red, cyan and magenta, respectively. Only 15 mHz and 100 mHz frequencies are shown to aid comparison to Figure \ref{fig:psumhisto} and visualization of the slopes.}
\label{fig:psds}
\end{figure}

To understand the influence of the origin of the power presented in Figure \ref{fig:psumhisto} on the slope of the power spectra, we compute the PSD for several averaging widths in Figure \ref{fig:psds}.
For $\alpha{=}2$ the contribution of power from larger angles is low and, therefore, the spectral energy distribution $P(f;\theta{=}0{-}10^\circ)$ at $\alpha{=}2$ is almost identical to $P(f;\theta{=}0{-}10^\circ)$ at $\alpha{=}1$ (see Fig. \ref{fig:psds}, green and blue). For $\alpha{=}50$ most of the coefficients are falsely associated with contributions from angles of $20^\circ$ and $30^\circ$ for 15~mHz and 100~mHz, respectively. This shows how large averaging widths smooth out local small scale variations and may thus lead to false angle association. However, as power from larger angles at increasing factors does not contribute equally to the $0{-}10^\circ$ bin for 15 mHz and 100 mHz, the slope appears to be similar. The amount of power of larger angles associated with the $\theta{=}0{-}10^\circ$ bin at $\alpha{=}10,50$ (Fig. \ref{fig:psumhisto}, red and cyan) for 15 mHz is much larger than for 100 mHz. From this follows that the spectral index can still be as steep as -2 and even be steeper than at $\alpha{=}1$, but the total power clearly increased. Due to this unequal contribution of power of larger angles at different frequencies, the spectral index can remain as steep as -2 to factors of $\alpha{\approx}20{-}50$ even though power is added to the parallel spectrum. For an increased averaging width by a factor of $\alpha{\geq}50$ the spectral index still shows anisotropy, but is more shallow than -2. For $\alpha{\geq}200$ the parallel spectral index is -5/3 and no spectral anisotropy can be resolved any more (see Figure \ref{fig:correlation_spec_factor}). Despite the fact that $\kappa$ stays around -2 for $\alpha$ up to 50, the magnetic field averaged at factors $\alpha{>}5$ should not be considered an appropriate local background magnetic field, as false coefficients contribute to the power.

\section{Discussion \& Conclusions}

We have introduced the angle uncertainty between the orientation of the averaged magnetic field and the orientation of the eddy fluctuations as a measure to describe the uncertainty of the orientation of a (local) background magnetic field and thus as a measure to resolve anisotropy of turbulent properties. We studied the scale dependent angle uncertainty and how it is related to the resolution of spectral anisotropy in fast solar wind data. 

The RMS of the angle uncertainty $\rmsdt$ depends on the frequency/eddy size of the fluctuation. A finite RMS implies the existence of a basic, frequency dependent, uncertainty to resolve anisotropy. We investigated previously observed anisotropy with a resolution of 10$^\circ$ \citep{Horbury2008}. Only if the RMS of the angle uncertainty is lower than 10$^\circ$ the correct association of magnetic field orientation to the wavelet coefficients can be assured. Based on the results presented in Figure \ref{fig:dtheta_factor}, such a correct association is obtained for averaging widths $\alpha\leq4$, which corresponds to $s_b{\leq}10s_e$. It is apparent that the definition of a local background magnetic field depends on the frequency range or eddy size under consideration and on the anisotropy to be resolved. Anisotropy sensitive to changes below 10$^\circ$ would require smaller averaging scales of the magnetic field to resolve such an anisotropy in case the distribution of scale dependent energy is similar to case studies here.

The solar wind observations presented in this work show that observed spectral anisotropy is not adequately resolved any more for $\alpha{\geq}50$ and vanishes for $\alpha{\geq}200$, i.e. an averaging width of more than 200 times larger than the eddy scale. Although the RMS of the angle uncertainty at factors $\alpha{\geq}5$ is larger than the width of the angle bin the spectral index remains anisotropic. This unexpected apparent robustness of the spectral anisotropy with respect to increased averaging width can be explained by a frequency dependent gain of power from wavelet coefficients of higher angles. Higher frequencies gain less power from wavelet coefficients associated with higher angles than lower frequencies (see Figure \ref{fig:psumhisto}). The origin of the power, meaning the angles associated with the power when averaging with $\alpha{=}1$, at different frequencies is in agreement with the frequency dependent RMS of the angle uncertainty. Even though the total power at small angles clearly increases for increasing factors $\alpha$ (see Figure \ref{fig:psds}), the slope of the PSD remains steep even for very large averaging widths. The apparent robustness to the increased averaging width should therefore not lead to an incorrect conclusion on the size of a local background magnetic field. The RMS of the angle uncertainty predicts the error in the association of wavelet coefficients and for $\alpha{\geq}5$ the total power in the parallel spectrum clearly increases. Only the anisotropy in the spectral index appears to be intact due to the effect of frequency dependent power gain.

Within the studied data set, a magnetic field averaged at $\alpha{\approx}5{-}50$, may be regarded as an intermediate background magnetic field, not local nor global. For an intermediate background magnetic field, although scale dependently averaged, the RMS of the angle uncertainty $\rmsdt$ within the frequency range under consideration is larger than the accuracy needed (10$^\circ$) and wavelet coefficients may not be linked to the correct angle. 

At even larger averaging scales corresponding to $\alpha{\geq}50$, the background magnetic field approaches the global mean magnetic field. This is often referred to as a magnetic field averaged scale independently \citep{Oughton2015}. Even if averaged scale dependently, a magnetic field on the order of the outer scale $L{\sim}1.5\cdot 10^6$ km may be regarded as the global mean magnetic field \citep{Wicks2010}. The RMS of the angle uncertainty $\rmsdt$ within the analyzed frequency range, exceeds 25$^\circ$ for averages over scales larger than the outer scale (see Figure \ref{fig:dtheta_factor}, dashed lines). In this case a significant amount of power cannot be linked correctly to a field to flow angle bin. 

For comparison between local and global background magnetic field we also analyzed the data using a scale-independent background field (not shown). Here, the power of eddies at different scales is associated with the same background magnetic field. We were unable to observe spectra which scale with $f^{-2}$ parallel to the background magnetic field using such a global frame. The RMS of the angle uncertainty of such a scale-independent magnetic field also depends on the frequency under consideration. When averaging with a window width of 5 times the largest period ($5\cdot67$ s) of the frequencies within which spectral indices are calculated, the RMS of the angle uncertainty ranges from 7$^\circ$ at low frequencies to 10$^\circ$ at high frequencies (see Figure \ref{fig:dtheta_factor} at $\alpha{\approx}2$ for 15 mHz and $\alpha{\approx} 13$ for 100 mHz). Eddies of lower frequency with periods closer to the averaging scale might be represented well enough by such a background magnetic field. However, as higher frequencies are analyzed with the same background magnetic field, the power of these eddies is associated with an angle resulting from larger scale fluctuations. Consequently more power from \emph{false} coefficients contribute to higher frequency fluctuations whereas very few (or none) contribute to the lower frequency fluctuations. Following this, the effect that the spectral index can remain steep, observed for scale dependent magnetic fields, does not hold for the global approach. Note, in the limit of a very strong background magnetic field, where $|\vec{B}_0|\gg |\vec{\delta b}|$, the values of $\delta\theta$ decrease and may have a negligible frequency dependency and the global approach will be applicable.

The angle uncertainty presented here provides an uncertainty measure of the orientation of turbulent structures/eddies as function of scale of the averaged magnetic field and of the associated frequency/period of the eddies under investigation. This method, however, does not constrain the scale of the wave numbers of the eddies in direction parallel and perpendicular to the associated background magnetic field. The reason is that under the assumption of Taylor's hypothesis \citep{Taylor1938} and spatial and temporal stationarity of the magnetic field, wave vectors of different magnitude and orientation contribute to the spectral energy density at one frequency $f_e$ (e.g., \citealt{Fredricks1976, Papen2015}).

The angle uncertainty is physically controlled by two effects: a) the frequency/scale dependent amplitudes of the turbulent fluctuations and b) any non-turbulent contributions such as the magnetic field convected out from the solar corona. These contributions generate the total field, which controls the orientation of the turbulent eddies. Thus the contribution of the solar background field \citep{Parker1958} with respect to the amplitude of the fluctuations plays an important role for the angle uncertainty. For example, for magnetic field fluctuations $\vec{\delta b}$ much smaller than the amplitude of the global mean magnetic field $\vec{B}_0$, the angle uncertainty would tend to small values and anisotropy should be well resolved. The ability to resolve anisotropy as a function of scale is not universally equal, but depends on the turbulent system, for example on the values of the spectral slopes of the energy distribution. The angle uncertainty is a helpful measure that can be applied to various systems to evaluate the ability to resolve anisotropy with a certain degree.




\listofchanges

\end{document}